\documentclass{emulateapj}

\def\etal{et al.\ }

\def\p3m{P${}^3$M}
\def\ap3m{AP${}^3$M}
\def\-{{\em{---}}}
\def\msun{{M_\odot}}

\newcommand{\be}{\begin{equation}}
\newcommand{\ba}{\begin{eqnarray}}
\newcommand{\ee}{\end{equation}}
\newcommand{\ea}{\end{eqnarray}}

\begin{document}
\title{The Spatial Distribution of the Galactic First Stars II: SPH
Approach} \author{Chris B. Brook$^{1,2}$, Daisuke Kawata$^{3,4}$,  Evan
Scannapieco$^{5}$, Hugo Martel$^{1}$, and Brad K. Gibson$^{6}$}
\altaffiltext{1}{D\'epartement de physique, de g\'enie physique et
d'optique, Universit\'e Laval, Qu\'ebec, QC, Canada  G1K 7P4}
\altaffiltext{2}{Dept. of Astronomy, University of Washington, Box 351580, Seattle, WA 98195, USA}
\altaffiltext{3}{The Observatories of the Carnegie Institution of
Washington,  813 Santa Barbara St., Pasadena, CA 91101}
\altaffiltext{4}{Swinburne University of Technology, Hawthorn VIC
3122, Australia} \altaffiltext{5}{Kavli Institute for Theoretical
Physics, Kohn Hall, UC Santa Barbara, Santa Barbara, CA 93106}
\altaffiltext{6}{Centre for Astrophysics, University of Central
Lancashire,  Preston, PR1 2HE, United Kingdom}

\begin{abstract}

We use cosmological, chemo-dynamical, smoothed particle hydrodynamical
simulations of Milky-Way-analogue galaxies to find the expected
present-day  distributions of both metal-free stars that formed
from primordial gas and the oldest star populations.  We find that
metal-free stars continue to form until $z\sim 4$ in halos that are
chemically isolated and located far away from the biggest progenitor
of the final system.  As a result, if the Population III initial mass function allows stars with low enough mass to survive until $z=0$ ($< 0.8 M_\odot$), they  would be  distributed throughout the Galactic halo. On the other hand, the oldest stars
form in halos that collapsed  close to the highest density peak of the
final system, and at $z=0$ they are located preferentially in the
central region of the Galaxy, i.e., in the  bulge. According to our models, these trends  are not sensitive to the merger
histories of the disk galaxies or the implementation of supernova feedback.
Furthermore, these full hydrodynamics results are consistent with our
$N$-body results in Paper I, and lend further weight to the conclusion
that surveys of low-metallicity stars in the Galactic halo can be
used to directly constrain the properties of primordial stars.  In
particular, they suggest that the current lack of detections  of
metal-free stars implies that their lifetimes were shorter than a
Hubble  time, placing constraints on the metal-free initial mass
function.

\end{abstract}


\section{Introduction}

The first generation of stars in the Universe  are expected to be
metal-free.  If some of these stars have lifetimes of   a Hubble time,
metal-free stars would be currently  found in the Milky Way.  Hunting
 metal-free stars, i.e., Population III (Pop III) stars, is an important
goal of contemporary astronomy, and  extensive searches for
extremely low-metallicity stars  are underway [e.g. The Hamburg/ESO 
R-process-Enhanced Star (HERES) survey, and the Sloan Extension for 
Galactic Understanding and Evolution (SEGUE); 
see \citealt{beers05} for an overview of past and current surveys].
To date,  surveys have been primarily in the solar neighborhood, which has the
obvious advantage over the bulge of proximity, as well as lacking the
dust extinction and crowding.  Searches also target the outer halo,
which retains the latter two advantages.  Currently, only 12 stars
with iron abundances $\rm [Fe/H]<-3.5$  have been identified
(\citealt{beers05}), with the lowest having $\rm [Fe/H]=-5.4$
(\citealt{frebel}).  Thus far, however, there is no report of the
detection of a single star that does not contain any metals.
Therefore, one crucial question is raised: is the current survey area
the right place to look for the metal-free stars?

Recently, \citet{diemand} used high-resolution N-body simulations to
trace the positions of the particles contained in the protogalactic
halos collapsing from $3.0\sigma$ and $3.5\sigma$  density
perturbations.  Tracing particles in $3.0\sigma$ perturbations, they
found that the density of such particles in the solar neighborhood is
three orders of magnitude lower than in the bulge at $z=0$.  Moreover,
tracing particles in $3.5\sigma$ perturbations leads to even more
extreme results, decreasing the number of these particles at the solar
radius by almost another order of magnitude.  This implies that 
the Milky Way's bulge is the best place to  search for the  very
oldest stars and their remnants. Therefore,  if Pop III and the oldest
stars are one and the same,  the non-detection of Pop III stars may
simply be due to the surveys  looking  in the wrong place,  and
future surveys should focus on the bulge region.

However, the link between the oldest stars, metal poor stars,  and  metal-free Pop III
stars is not so straight forward. The  largest halos, forming in the
greatest density perturbations, can progress their self-enrichment,
and produce the next generation of stars, at very  high
redshift.   Combining semi-analytic galaxy formation recipes  of
\cite{kauf} with N-body simulations of galaxy clusters, rescaled  to
Milky Way mass, \cite{white} showed that  the oldest stars are
distributed preferentially at small radii, while low metallicity stars
(simply assumed to form in the lowest mass  halos) lie preferentially
at larger radii in the present Milky Way. This  effect is driven by
the flatter density profile of low $\sigma$ peaks,  compared to high
$\sigma$ peaks as shown, for example, at $z=12$ in figure~2 of
\cite{Moore}, and assuming the well established galaxy mass-metallicity relation. However, the association of low metallicity stars with low mass halos does not necessarly mean that the earliest stars in these halos formed from primordial material. The low metallicity of dwarf galaxies may be due to factors 
caused by inefficient star formation \cite{kennicut}, feedback driven outflow of the enriched gas, because of  their shallow potential well \cite{dekel}, or inflow of low metallicity material \cite{edmund}. This may occur even if the earliest stars in such halos form from (slightly) enriched gas. But these studies [see also an
analytical argument by \citet{escude}] do raise the possibility that  Pop III stars have a
different radial distribution from the oldest stars, i.e. it is possible that low-density, isolated
regions may form stars from primordial material at (relatively) low
redshift.

To identify the best location to look for Pop III stars, it is
necessary  to infuse the dark matter structure formation studies employed in  \cite{white} and \cite{diemand}
with information pertaining to the metal  enrichment of the intergalactic medium.  In \citet{evan06}
(hereafter Paper I), we combined a high-resolution N-body simulation
of the formation of the Milky Way with a semi-analytic model of
metal enrichment.  This model traced the outflows induced by star
formation in the collapsing halos identified in the N-body simulation,
and  tagged the metal-free  star particles that were born in halos massive
enough to make stars, but isolated enough to not have been enriched by
outflows from neighboring halos. We found that Pop III stars were
formed over a large redshift range, that peaked at $z\approx 10$, but
continued to form down to $z\approx 5$. Tracing these particles to their
final locations at $z=0$,  we found that the Pop III distribution
covered a wide range of galactocentric radii.

A limitation of this approach, however, is the assumption used for
tracing baryonic  matter, with the density profile of stellar matter
assumed to be the same as  that for the dark matter. In this paper
(Paper II), we use a different  approach: chemo-dynamical simulations.
Continued advances in computer technology and numerical
methods have made it possible to calculate the dynamical and chemical
evolution of disk galaxies self-consistently
(e.g. \citealt{bc01,abadi,nn03,brook04,sm05,stws05}). Such
chemo-dynamical simulations can tell us the location of stars with
different metallicities and ages at different redshifts. In this
paper, we use such information from cosmological chemo-dynamical
simulations of Milky-Way analogues to study the birthplace and final
distribution of the zero-metallicity stars as well as very old stars.
While working at a somewhat lower resolution than in Paper I, this
approach nevertheless has the advantage of self-consistently tracing
the details of star formation and metal enrichment within dynamical,
hierarchical structure formation. In particular, we examine four
simulations of Milky-Way-type galaxies which employ various
implementations of supernovae (SNe) feedback, different resolutions,
and different random initial density fluctuations. Interestingly, we find that the resulting distributions of the oldest and zero-metallicity stars are consistent among the four simulations, 
robust within our different models.

In \S2 we introduce the code and models which we employ.  In \S3 we
show the star formation histories of the zero-metallicity ``metal-free''
stars. We then show the distribution of such stars, as well as the
distribution of very ``oldest'' stars, the earliest stars forming in
each simulation.
A summary of our results and our conclusions follow in \S4.

\section{Method} 

\subsection{The Disk Galaxy Simulation Models}
We simulate four Milky-Way-analogue galaxies using the chemo-dynamical
software GCD+, which self-consistently models
the effects of gravity, gas dynamics, radiative cooling, and star
formation (\citealt{kawataa}).
Two of the models, AFM and TFM, 
have identical initial conditions but
different prescriptions for modeling the SNe feedback.
The other two models, KGCD and AGCD, are higher resolution
disk galaxy formation models used in \citet{bailin}. 
We apply the same analysis for all four models
with varying evolution histories,
prescriptions of star formation and supernovae (SNe) feedback,
in order to test the robustness of our conclusions.

  All models adopt $\Lambda$-dominated CDM cosmological
simulations that use the multi-mass technique to 
self-consistently model the large-scale
tidal field, while simulating the galactic disk at high resolution.
The initial conditions used for AFM and TFM corresponds to galaxy D1 of
Kawata, Gibson, \& Windhorst (2004), but in this simulation,
we set the high-resolution region to
be 8 times the virial radius of this galaxy at $z=0$.
The initial condition of AFM, TFM, and KGCD are constructed with 
{\tt GRAFIC2} (Bertschinger 2001)
and that of AGCD is the same as \citet{abadi} which 
is kindly provided by M. Steinmetz. All initial conditions have been shown 
to lead to disk  galaxies.
Detail description about these models is
seen in these earlier papers. Here, we provide a brief description of
these models.

  In TFM, KGCD, and AGCD, we employ the exactly same recipe as
\citet{kawataa}. In these models, $10^{50}$ erg energy is 
fed back as thermal energy from each SN, and
energy from SNe II and SNe Ia is smoothed over
surrounding gas particles according to SPH kernel, in
the form of thermal energy, as in \citet{kats}.
This SNe feedback model is often called thermal feedback model.
On the other hand, AFM adopts the adiabatic
feedback model introduced in \citet{brook04}. 
In this model, to maximize the effect of SNe feedback,
we assume $10^{51}$ erg energy per SN.
In addition, gas within the SPH smoothing kernel of SNe II
explosions is prevented from cooling. This adiabatic phase
is assumed to last for the lifetime of the lowest mass star
that ends as a SN II, i.e., the lifetime of 8 $M_\odot$ star 
($\sim100$ Myr). The energy released by SNe Ia, which do
not so closely trace the starburst region, is not assumed to have an adiabatic 
phase.

  Thus, with our four models we have three different random seed initial 
conditions, with two different feedback formalisms, and run at three 
different resolutions. Table \ref{tab:params} summarizes the parameters 
of all the
simulations. All the models assume the Salpeter initial mass function (IMF)
with the mass range of $0.1-60$ $M_{\odot}$.
If the IMF for early generation stars are different from the conventional Salpeter IMF, with more massive stars as suggested in the numerical simluations, the total effect
of SNe feedback may be stronger than in our simulations. This may reduce the number of the old and/or metal-free stars. Therefore, the absolute mass density of these stars in
our simulations needs to be taken with caution. However,  the main aim of this study is to identify the best place to look for the metal-free stars, if they still exist. Thus the important result we want to highlight is the radial distribution rather than the absolute density of the metal free stars. This aspect of our result will only be invalidated if the increased feedback were to greatly effect the enrichment of the IGM, such that isolated halos form from pre-enriched rather than primordial material. In the parameter space survey in Paper I, the strength of that approach was its ability to probe this aspect of the problem for a large range of feedback energies, assuming a variety of top heavy IMFs, and this gives us confidence. Under these assumptions, we can study the expected distribution of the
  the metal-free stars in the Milky Way at $z=0$, if they had a similar
 IMF as the local stars.

\begin{deluxetable*}{lrrrrrrrrrrr}
\tablecaption{
 Parameters of the simulations
}
\tablewidth{0pt}
\tablehead{\colhead{Model} & \colhead{$M_{\rm vir}$} &
 \colhead{$r_{\rm vir}$} & \colhead{$m_{\rm gas}$\tablenotemark{a}} & 
 \colhead{$m_{\rm DM}$\tablenotemark{b}}  &
 \colhead{$\epsilon_{\rm gas}$\tablenotemark{c}} &
 \colhead{$\epsilon_{\rm DM}$\tablenotemark{d}} & 
 \colhead{$\Omega_0$} & \colhead{$\Lambda_0$} &  \colhead{$h_0$}  &
 \colhead{$\Omega_b$}  &  \colhead{$\sigma_8$} \\
                 & \colhead{($M_{\odot}$)} & 
 \colhead{(kpc)} & \colhead{($M_{\odot}$)} & \colhead{($M_{\odot}$)} &
 \colhead{(kpc)}   & \colhead{(kpc)}         & &  &
                 &                         &   \\
  }
\startdata
AFM & 10.$\times10^{11}$ & 275  & 7.3$\times10^6$  & 
 4.9$\times10^7$ & 0.93 & 1.6  & 0.3 & 0.7 & 0.7 & 0.039 & 0.9  \cr
TFM & 4.9$\times10^{11}$ & 275 & 7.3$\times10^6$  & 
 4.9$\times10^7$ & 0.93 & 1.6 & 0.3 & 0.7 & 0.7 & 0.039 & 0.9 \cr
KGCD & 8.8$\times10^{11}$ & 240 & 9.2$\times10^5$ 
 & 6.2$\times10^6$ & 0.57 & 1.1 & 0.3 & 
  0.7 & 0.7 & 0.039 & 0.9 \cr
 AGCD & 9.3$\times10^{11}$ & 270 & 3.3$\times10^6$ & 
 1.9$\times10^7$ & 0.87 & 1.5 & 0.3 & 0.7 & 0.65 & 0.045 & 0.9 \cr
\enddata
\label{tab:params}
\tablenotetext{a}{Mass of gas particles.}
\tablenotetext{b}{Mass of dark matter particles.}
\tablenotetext{c}{Softening length of gas particles.}
\tablenotetext{d}{Softening length of dark matter particles.}
\end{deluxetable*}

\subsection{Identifying Metal-Free Stars}

In Paper I, we combined a high-resolution N-body simulation that can
resolve the formation of halos $\sim10^7M_{\odot}$ with a
semi-analytic model of outflows.  In our semi-analytic model, we
considered that the outflows are induced by star formation in the
halos, depending on their total mass, and traced the enrichment
history of  the intergalactic medium due to such outflows. Then, we
identified the site of metal-free star formation as halos which are massive
enough to make stars and have not been enriched by the outflows from
other halos.

It is also worth mentioning that numerical simulations have
    shown that outflows tend to be highly anisotropic, even if
    they originate from spherical halos  \citep{ms01a,ms01b}.
    The external medium surrounding halos tends to be
    anisotropic. When gas ejected from halos enters that medium,
    it naturally follows the path of least resistance, becoming
    anisotropic. Energy and metals are deposited in low-density
    regions, away from dense structures.
 This effect strongly reduces the
likelihood of galaxies being hit by outflows originating from other
galaxies, unless the outflows are very narrow and can travel large
distances, across cosmological voids \citep{pmg06}. The net effect is
a significant suppression of the metal-enrichment of halos by
neighboring galaxies, enabling these halos to retain their primordial
composition up to low redshifts, when they may form stars.

The chemo-dynamical simulations employed in this present study do
not require the inclusion of an approximate semi-analytic model.
Rather, they include the effects of the SNe feedback completely
self-consistently within the SPH method. First-generation star
particles form on an individual basis, and we track the independent
chemical evolution of each and every gas particle, starting from
primordial gas. We then use these individual metal-free star
particles as tracers of metal-free stars, and analyze their locations at
$z=0$. 

It is believed that the bulk of  metal-free stars are
  formed  in  halos with total mass around $\sim10^7M_\odot$
  (\citealt{tegmark,fuller,yoshida}).  
      However, it is unknown what
  fraction of the gas in the halo turns into metal-free stars
  (e.g. \citealt{abel00}), whether the SNe by metal-free stars disrupt the gas
  in the halo completely \citep{bromm03}, although full radiative transfer calculations of {\cite{kitayama} indicate that  all the gas is evacuated,   or whether such SNe also induce
  the second generation stars in the halo \citep{susa}.  It is also possible that two types of metal free stars form, Pop III stars, with mass of  100M$_\odot$ and more, and metal free stars forming in halos with T$>10^4$K, with masses as low as 10M$_\odot$    \citep{bromm06}.
  Simulations of these processes have far greater resolution than we can manage when simulating a galaxy in  a cosmological context.  

Unfortunately, the resolution of cosmological, SPH galaxy formation simulations is not sufficient to identify the halos where the metal-free stars are likely
to form, or to follow the physical processes 
in the halos, in the manner of e.g. \cite{yosh06}. 
Rather, the minimum mass of the halo that we can resolve is
between $10^8 \msun$ and $10^9 \msun,$ depending on the simulation.
However, in a hierarchical clustering scenario, each and every $10^8
\msun - 10^9 \msun$ halo is expected to be pre-dated by 
smaller halos, i.e. building blocks.
Thus, if a given   $\approx 10^8-10^9 M_{\odot}$ halo is sufficiently
isolated,  its earliest forming smaller progenitor, such as
$\sim 10^7$ M$_\odot$ halos, should have been the site of metal-free star formation. 
Based on this scenario, we assume that some fraction of the mass of
simulation metal-free star particles is metal-free stars, and that the metal free simulation star particle trace the location of the metal-free stars. Thus our study assumes that the final distribution of stars in the stellar halo is primarily determined by the dynamics and accretion of the $\sim 10^8$M$_\odot$ halos that our simulations resolve. In this case the remaining uncertainty is how many ``metal-free
stars" correspond to each zero metallicity star particle at $z=0$.
As mentioned above, this depends on the star formation history 
in the $10^6-10^7$M$_\odot$ mini-halos which we do no  resolve. We simply assume
an upper limit, by assuming all the stars in that simulation star particle are metal-free,
and follow the Salpeter IMF. This assumption
guarantees that some fraction of metal-free stars survive
till z=0, and allows us to use the metal-free simulation star particles as 
tracers of the positon of metal-free stars. 
Therefore, our absolute mass density of the metal-free stars are
likely overestimated.  However, we are able to determine  the radial distribution of the metal-free stars, using this simple approach. This allows us to determine whether local observations can be used as a constraint on the Population III IMF.

Finally, we also trace the position of old stars, which we define as
the oldest star particles, regardless of whether they form from
primordial material. In order to compare their distributions with
metal-free stars, the mass of oldest and metal-free stars that we
trace are approximately equal, i.e. the total mass of the oldest star
particles and metal-free star particles are the same. As mentioned
above, we employ four disk galaxy formation models, that adopt
different galaxy formation models and different prescriptions of SNe
feedback. These test how the final distribution of metal-free
stars and the oldest stars are sensitive to the evolution history of
galaxies and the effect of the SNe feedback. Hereafter, we call the
simulation star particles that house metal-free  stars, i.e.\ first-generation
Pop III stars, and an equivalent mass of the very
oldest star particles as ``oldest'' stars.

The strengths and shortcomings of the SPH approach, which includes gas 
dynamics, cooling, star formation and feedback self consistently (Paper II), 
and the higher resolution N-body + semi-analytic  approach, which allows a 
greater diversity  of feedback implementations (Paper I) make the two studies 
complimentary. Reading the two  together allows us to gain greater confidence
in our results. 

\section{Results}

 In Figure \ref{sfr}, we plot the star formation rate (SFR) as a
function of redshift for our four simulated  Milky-Way analogues.  The
SFR for all stars within the virial radius at $z=0$ is plotted as a
dot-dashed line, and the SFR for metal-free stars as a solid line.  It
is notable that the AFM and TFM models, which have the same initial
conditions but different treatments of SNe feedback, have
significantly different star formation histories, with star formation
relatively delayed in the AFM run \citep[see][for details]{brook04}.
 
 Metal-free stars continue forming over a wide range of redshifts, from
before redshift $z=11$ down beyond redshift $z=4$. Even though it had
a relatively large effect on the star formation history of the galaxy,
the different feedback implemented in the AFM and TFM runs have not
resulted in a drastic change in the star formation history of metal-free stars. Although the AFM, KGCD, and AGCD form very few metal-free stars
after redshift $z=4$,  metal-free  stars do continue to form  beyond redshift
$z=3$.  Overall, the redshift range of met-free star formation is quite
consistent among the four models, starting along with the oldest stars
beyond redshift $z=11$ and continuing to $z\sim3$.  Furthermore, this
redshift range is roughly consistent with the weak  and intermediate
wind models in Paper I. As expected with the differences
between the models  explained above,  the higher resolution
simulations in Paper I allow some metal-free  stars to form at higher
redshift, $z\sim15$.

 Figures \ref{AFM}--\ref{AGCD} present the final distributions
of both metal-free  and oldest stars.
The top left panels show the distribution of the oldest 
(red ``$\times$'') and metal-free  (black ``+'') stars 
in X-Z plane, with axis of 100 kpc, and
where Z is the direction of the rotation axis of the stellar
disk at $z=0$. 
The upper right panels show the mass density of the two populations 
as a function of radius (distance from the center), 
in units of $M_\odot{\rm /pc^3}$. The stellar halo 
mass of each simulation is normalized to that of the Milky Way, by requiring that the solar neighborhood density matches that for the Milky Way, $5.7\times 10^5$ $M_\odot$pc$^{-3}$ (\citealt{preston}).

The lower left 
panel of figures~ \ref{AFM}--\ref{AGCD} show the radial dependence of the mass 
fraction of metal-free  and oldest stars with respect to
the total stellar mass at each radius at $z=0$.
The lower right panel displays the distributions
of the radius of the birth-place for metal-free  stars and oldest stars, which
is measured from the center of the largest
proto-galaxy at the time they born.

Although the details of the distribution of metal-free  and oldest stars 
differ among the four models,  several trends can be
seen to be common to all  models.
Not surprisingly, the absolute density of metal-free  and oldest stars 
are higher at smaller radii.
The oldest stars have steeper density profile than metal-free  stars, 
i.e. oldest stars are more centrally concentrated. 
In terms of where these stars are best searched for
in current surveys of stars, the most important quantity is not
the absolute density, but rather the number fraction,
with respect to the field stars.
This is shown in the lower left panels of Figures \ref{AFM}--\ref{AGCD}.
Here, we define the field stars as all stellar halo stars at that 
radius, computed by using the Toomre diagram and requiring 
that $T^2+V^2>(V-V_{\rm disk})^2$ where $T^2=U^2+W^2$, $U$, $V$, $W$ 
are the radial, rotational, and out-of-plane velocity
components, respectively, and $V_{\rm disk}$ is the 
rotation velocity of the stellar disk component. Here, $V_{\rm disk}$  for each model is measured as the mean rotation velocity for
  the young disk  stars (which formed after $z\sim0.8$)

These panels show clearly that the 
oldest stars are highly concentrated 
in the bulge regions in the center of the galaxies.
On the other hand, metal-free  stars have a 
 higher fraction in the outer regions, i.e. in  halo region.
This strong difference between the oldest and the metal-free 
stars is in fact more extreme than what was seen in
Paper I (see Figs. 4, 5, and 7 in that paper). 
In Paper I, our N-body$+$semi-analytic model
predicted a roughly constant fraction independent of the radius,
while the chemo-dynamical simulation model shown here predicts
that the outer region has a higher fraction of  metal-free  stars 
than the inner region. This is because the chemo-dynamical simulations naturally
 provide the density profile of the field stellar component, which
is more centrally concentrated than the dark matter component.
This difference leads to a difference in the profiles of the normalized
mass fraction of metal-free  stars in our two papers.
Nevertheless, if anything our conclusion from
the chemo-dynamical models is even stronger than that reached in Paper I, and
  we predict that the current survey area, i.e. the solar neighborhood
  and outer halo, is the best place to look for Pop III stars. Thus, from a completely different perspective,
we again find that {\it  if they have sufficiently long lifetimes, a
significant number of stars formed in initially primordial gas 
should be found in the Galactic halo.}

\begin{figure*}
\plotone{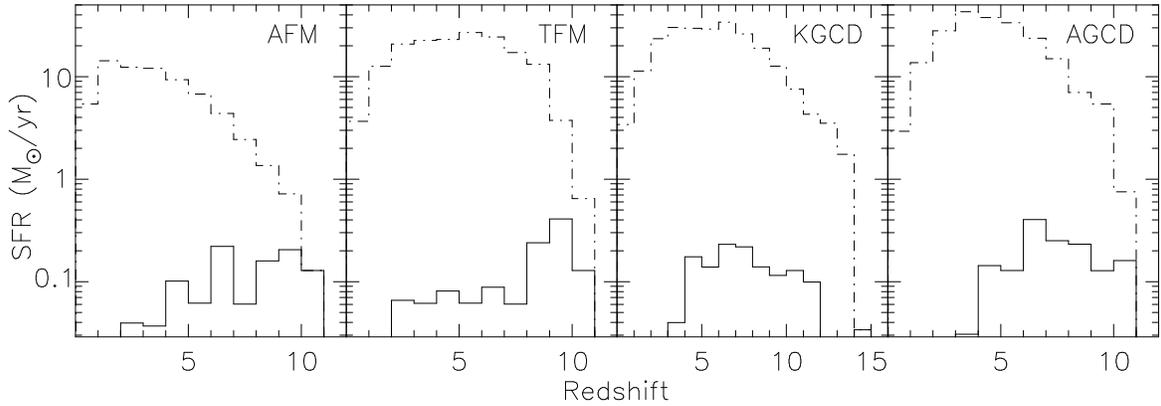}
\caption{Star formation rates  (SFR) in $M_\odot/{\rm yr}$ for the 4 models, from
   left to  right: AFM, TFM, KGCD, and AGCD. The dot-dashed line 
  shows all
  stars, while the solid line shows the SFR for stars formed from primordial material, or ``metal-free" stars.  
}
\label{sfr}
\end{figure*}

\begin{figure}
\plotone{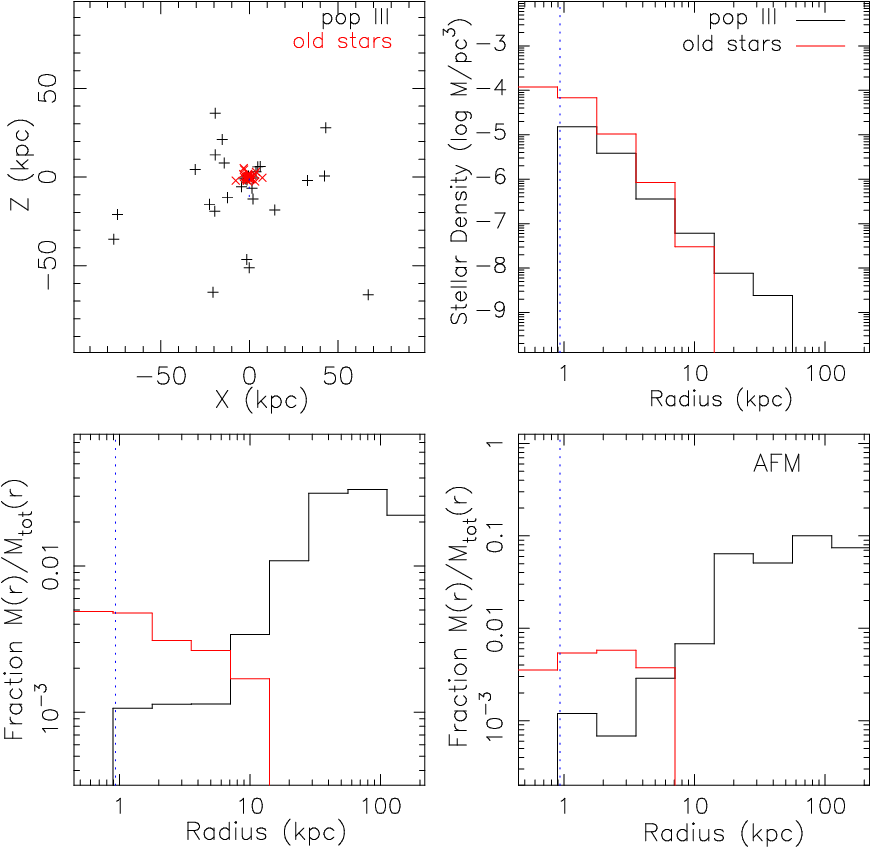}
\caption{The distribution of the oldest (red ``$\times$'') and metal-free (black ``+'') stars in the AFM are plotted in X-Z plane in
the top left panel, with axis of 100 kpc, and where Z is the
direction of the angular momentum vector. The oldest stars are
centrally concentrated in the bulge region, while the metal-free 
stars are spread throughout the halo. In the last three
panels, red lines show the oldest stars while black lines
are metal-free stars. The blue dotted line indicates the spatial resolution limitation of the simulations.  The upper right panel shows the mass
density of the two populations as a function of radius, in units of
$M_\odot/{\rm pc}^3$. The stellar halo density of the simulation is normalized 
to that of the Milky Way in the solar neighborhood.  Stellar mass 
loss from a top heavy IMF is also not accounted for. The lower left panel shows the radial
dependence of
the mass fraction of stars at $z=0$, for the two populations. The
lower right panel shows the radial dependence of the mass fraction
of the two populations of stars at their birth-place and
birth-time. For this plot, the center at each time is taken as the
center of the largest protogalaxy at that time. 
}
\label{AFM}
\end{figure}

\begin{figure}
\plotone{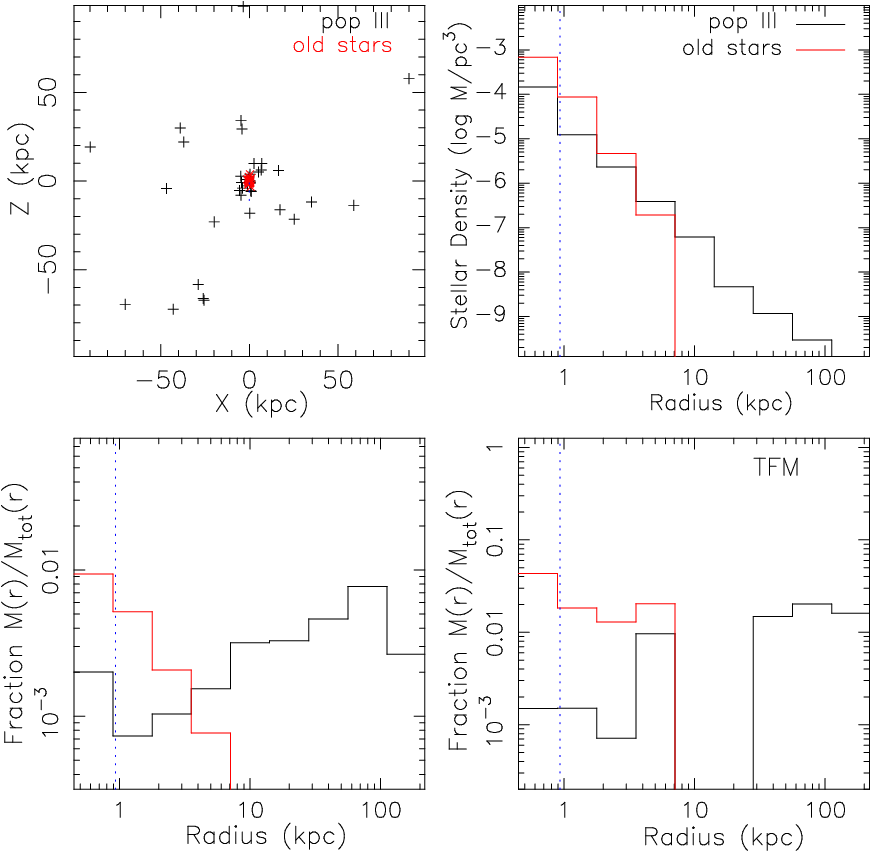}
\caption{ Same as Figure \ref{AFM}, for the TFM. No significant difference 
in trends are apparent, despite the difference in feedback models and 
subsequent difference in star formation histories .   
}
\label{TFM}
\end{figure}

\begin{figure}
\plotone{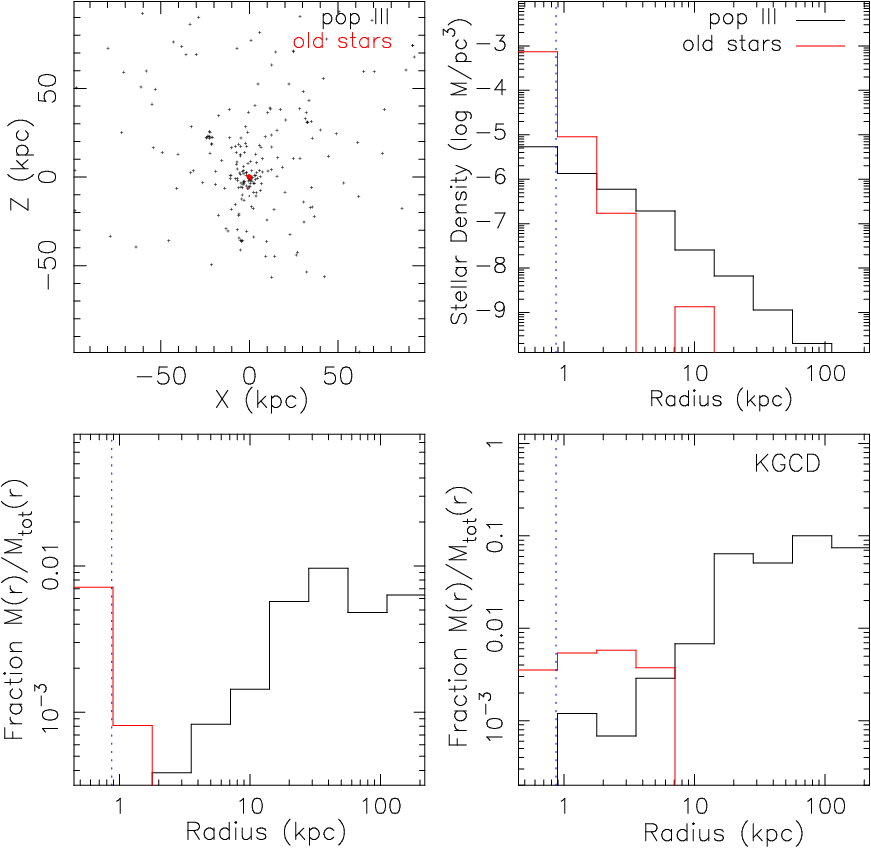}
\caption{ Distributions for KGCD model, using the formats of the
  previous two figures. 
}
\label{KGCD}
\end{figure}

\begin{figure}
\plotone{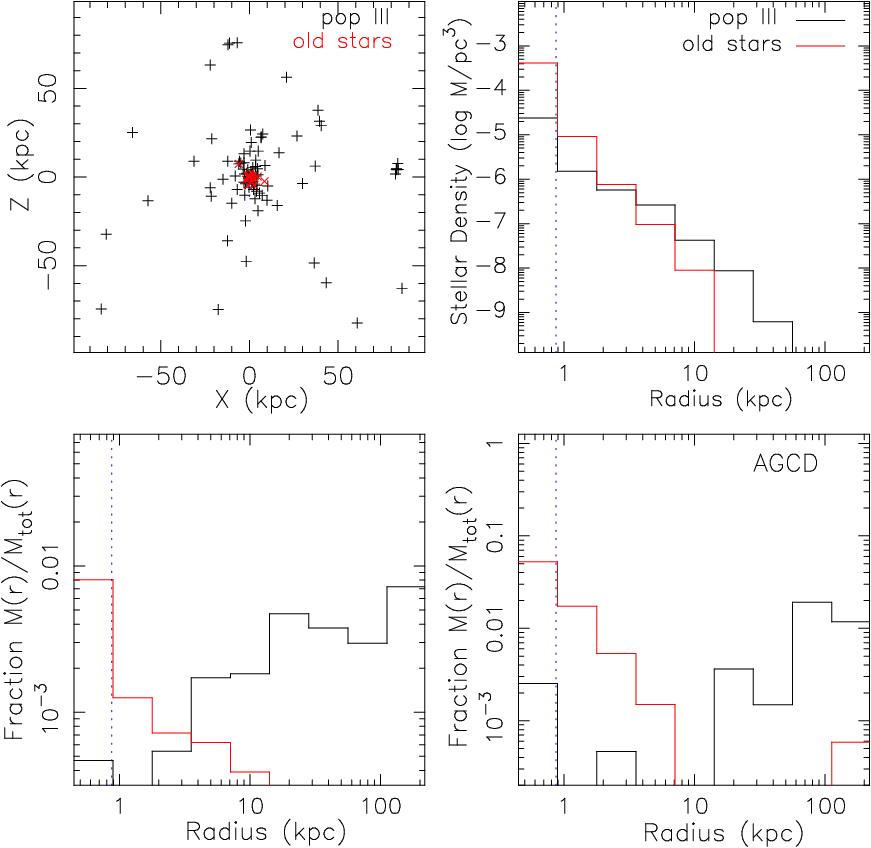}
\caption{ Distributions for AGCD model, using the formats of the
  previous three figures.
}
\label{AGCD}
\end{figure}

\begin{deluxetable}{lrr}
\tablecaption{
Average baryonic mass of the proto-galaxies where metal-free  
stars and oldest stars form (in units of $10^9\msun$).
}
\tablewidth{0pt}
\tablehead{\colhead{Model} & 
\colhead{metal-free  stars} & \colhead{oldest stars}
}
 \startdata
AFM  & 1.5 & 6.5 \\
TFM  & 1.3 & 6.8 \\
KGCD & 0.7 & 8.2 \\
AGCD & 2.1 & 4.3 \\
\enddata
\label{table}
\end{deluxetable}

While this overall trend is constant across our SPH models,
there are nevertheless a number of interesting differences
between the runs, which we now discuss in further detail.
First, we compare the TFM and AFM runs, i.e., the runs with
different implementations of SNe feedback.
Figures \ref{AFM} and \ref{TFM} show that 
both metal-free  and oldest stars are slightly  more centrally concentrated in 
TFM model than the AFM model. This is related to
their star formation histories in Figure \ref{sfr}.  
In TFM, a larger number of stars are 
born at very high redshifts, because there is
no mechanism to suppress star formation in the small 
proto-galaxies. As a result, oldest stars in the TFM run are an older population 
than those in the AFM run, and there is  
a larger fraction of metal-free  stars which are oldest stars, 
``metal-free$+$oldest stars.'' Due to the so-called bias effect, the halos near the 
density peak
of the final system form earlier, and the metal-free$+$oldest stars
form in these halos.
The lower-right panels of Figures \ref{AFM} and \ref{TFM}
demonstrate this clearly. In TFM, more metal-free  stars form
close to the center of the biggest galaxy, which leads to
the bimodal distribution of the distance of their birth place.
The stars which formed close to the center of the biggest galaxy 
accreted to the system at earlier epoch.
This leads to a larger number of metal-free $+$oldest stars ending up 
in the central region in TFM than AFM.

KGCD has the most clear difference in distributions between metal-free  and oldest 
stars, in terms of their mass fraction at different radii, with the model 
AGCD also showing a distinct difference in such distributions. In the
bottom right
panels of Figures \ref{AFM}--\ref{AGCD}, the largest progenitor is taken as 
being at the lowest radius. Examining the AGCD model for example, this diagram 
highlights how a relatively small number of metal-free  stars can pollute the 
progenitor in which  many of the oldest stars form. Yet at large distances from 
this largest, central progenitor, metal-free  stars continue to form.
 
 Three different initial conditions having three different
patterns of small-scale density perturbations have been employed in 
our models, which lead to
different merger histories of their progenitor halos, i.e., building blocks. 
Yet all models have similar spreads in the distribution of oldest stars, with a cutoff of the oldest stars around 10 kpc.
Lower panels of Figures \ref{AFM}--\ref{AGCD}
show how the final distribution of  metal-free  and oldest stars
reflects their birth places.  The oldest stars generally form in the 
largest progenitor, associated with the largest density peak, collapsing at the 
highest redshift, so that they tend to be in the inner region
  of the galaxy at $z=0$. Even though the final distributions of metal-free  and old 
stars depend to a degree on their merger histories, all the models lead to 
similar trends for the distributions of metal-free  and oldest stars.

 Although our chemo-dynamical simulations are not capable of
resolving the minimum mass of halos which can form 
metal-free  stars, $\sim10^7$ $M_{\odot}$, it is still interesting
to see the mass of the halo in which metal-free  and oldest stars formed.
Using a friend of friends (FOF) algorithm at each time-step, we are
able to calculate the mass of the halos in which each star is
born. We are limited here by time resolution of the simulation outputs, 
where we need to assume that the mass of an object has not changed 
significantly between the output time steps, $\sim0.2$ Gyr.
Table~\ref{table} presents the mean total masses of halos
in which metal-free  and oldest stars formed, respectively, where we define the
total mass of the halo as a total (dark matter 
and baryon) mass within the halo 
found by our FOF algorithm. It is clear that the masses of the halos 
where metal-free  
stars form are on average lower than those of halos where oldest stars form. 
This ranges from a factor of $\sim2$ larger in the
AGCD simulation to a factor of  $\sim 12$ in the higher-resolution KGCD 
simulation.
Combined with the radial distribution of the birth-places of the metal-free
stars, shown in Figures \ref{AFM}--\ref{AGCD}, it is clear
that metal-free  stars form in lower mass halos that
collapse at later epochs, but are chemically isolated from the more massive,
earlier collapsing halos in which the oldest stars are born. 
Consequently, metal-free  stars have a much greater spread in distribution,
compared with oldest stars.

\section{Summary and Conclusions}

Using chemo-dynamical simulations of Milky-Way-analogue galaxies, we study the 
present day distribution of stars formed from primordial material, i.e., 
first-generation stars (``metal-free  stars"), and an equivalent mass of the very 
earliest stars formed (``oldest stars").
Our simulations employ various implementations of energy feedback, 
different resolutions, and different random initial density
fluctuations. We find consistent trends in the distributions of 
the metal-free  and oldest stars in our four diverse simulations
at least withing the limitations of our modeling. The main results are as follows:

\begin{itemize}

\item Metal-free  stars can form in isolated proto-galaxies until relatively low 
redshift. Therefore, metal-free  stars are not necessarily among the very oldest 
stars.

\item Metal-free  stars have different radial distribution profiles
from oldest stars. The oldest stars have a more centrally concentrated
distribution, i.e., they are preferentially found in the bulge regions
of the Milky Way. On the other hand, metal-free  stars 
are distributed through the halo, 
and in terms of the fraction with respect to the field stars,
metal-free  stars are easier to be found in the outer regions. 

\end{itemize}
In addition, our chemodynamical simulation approach suggests:

\begin{itemize}
 \item The stellar halo density profile has a steeper slope than
   the dark matter density profile. As a result, the mass fraction
   of the metal-free stars with respect to the halo stars
   are higher in the outer region. The almost flat radial trend
   found in Paper I is due to the assumption that the stellar density
   profile is the same as the dark matter density profile.
\end{itemize}

The results are driven by  the combination of two results: (1) the contribution of
different sigma overdensities have different density profiles in the
final halo, ie lower sigma peaks have flatter density profiles (Moore
et al 2006), and the fact that the isolated, relatively late forming proto-galaxies (i.e. those forming in low sigma peaks), form from primordial material.

In terms our question ``where are the metal-free stars?'', our
chemo-dynamical simulation approach has reached 
the same answer as Paper I. Thus, we repeat here our conclusion
in Paper I:
if they have sufficiently long lifetimes, a significant number of stars
formed in initially primordial star clusters should be found in the
Galactic halo. This means that there is no compelling theoretical
reason to motivate observational searches in more difficult
environments, i.e., in the bulge region,
and present observations should be taken as directly
constraining the distribution of Population III (Pop III) stars.
The lack of metal-free halo stars today should be taken as strong
evidence of a $M\geq0.8$ $M_{\odot}$ lower limit on the 
IMF of metal-free stars.
Our results provide encouragement for observers of extreme low-metallicity 
stars in the solar neighborhood, which have direct implications on our 
knowledge of the chemical properties of  Pop III stars, 
and encourage theorists who wish to use such results as the
basis of their analysis of the nature of the metal-free stars. 

\acknowledgements

We thank M. Pieri for helpful comments and useful
information.  We acknowledge the Astronomical Data Analysis Center of
the National Astronomical Observatory, Japan (project ID: wmn14a), the
Institute of Space and Astronautical Science of Japan Aerospace
Exploration Agency, the Australian and Victorian Partnerships for
Advanced Computing, and the Laboratoire d'Astrophysique
Num\'erique, Universit\'e Laval. CB and HM are supported by 
the Canada Research Chair program and NSERC.
DK thanks the JSPS for financial support
through a Postdoctoral Fellowship for research abroad.
ES is supported by the National Science Foundation
under grant PHY99-07949. 

\fontsize{10}{10pt}\selectfont


\begin{thebibliography}{99}

\bibitem[Abadi et al.(2003)]{abadi}
 Abadi, M. G., Navarro, J. F., Steinmetz, M., \& Eke, V. R. 2003, 
\apj, 591, 499

\bibitem[Abel, Bryan, \& Norman(2000)]{abel00}
 Abel, T., Bryan, G. L., \& Norman, M. L. 2000, \apj, 540, 39

\bibitem[Bailin et al.(2005)]{bailin} 
Bailin, J. \etal 2005, \apj, 627, L17

\bibitem[Beers \& Christlieb(2005)]{beers05}
Beers, T. C., \& Christlieb, N. 2005, ARA\&A, 43, 531

\bibitem[Bekki \& Chiba(2001)]{bc01}
 Bekki, K., \& Chiba, M.\ 2001, \apj, 558, 666

\bibitem[Bromm, Yoshida, \& Hernquist (2003)]{bromm03}
Bromm, V., Yoshida, Y., \& Hernquist, L. 2003, \apj, 596, L135

\bibitem[Brook et al.(2004)]{brook04}
Brook, C. B., Kawata, D., Gibson, B. K., \& Flynn, C. 2004, \mnras, 349,
52

\bibitem[Dekel \& Silk (1986)]{dekel}
Dekel, A. \& Silk, J. 1986 ApJ, 303, 39

\bibitem[Diemand, Madau, \& Moore(2005)]{diemand} 
Diemand, J., Madau, P., \& Moore, B. 2005, \mnras, 364, 367

\bibitem[(K\"{o}ppen \& Edmunds (1999)]{edmund}
K\"{o}ppen \& Edmunds 1999 MNRAS, 306, 317

\bibitem[Frebel et al.(2005)]{frebel}
 Frebel, A. et al. 2005, Nature, 434, 871

\bibitem[Fuller \& Couchman(2000)]{fuller}
Fuller, T. M., \& Couchman, H. M. P. 2000, \apj,  544, 6

\bibitem[Greif \& Bromm(2006)]{bromm06}
Greif, T. H., \& Bromm, V. 2006, submitted to MNRAS (astro-ph/0604367) 

\bibitem[Katz(1992)]{kats}
Katz, N. 1992, ApJ, 391, 502


\bibitem[Kawata \& Gibson(2003)]{kawataa} 
Kawata, D., \& Gibson, B. K. 2003, \mnras, 340, 908

\bibitem[Kauffmann et al.(1999)]{kauf}
Kauffmann, G., Colberg, J. G., Diaferio, A., \& White, S. D. M. 1999, 
\mnras, 303, 188

\bibitem[Kennicut (1998)]{kennicut}
Kennicut, R. C., 1998 ApJ 498, 541

\bibitem[Kitayama \& Yoshida (2005)]{kitayama}
Kitayama, T. \& Yoshida, N.  2005, ApJ, 630, 675


\bibitem[Martel \& Shapiro(2001a)]{ms01a}
 Martel, H., \& Shapiro, P. R.\ 2001a, Rev.Mex.A\&A(SC), 10, 101

\bibitem[Martel \& Shapiro(2001b)]{ms01b}
 Martel, H., \& Shapiro, P. R.\ 2001b, in Relativistic Astrophysics,
AIP Conference Proceedings 586, eds. J. C. Wheeler \& H. Martel, p.~265

\bibitem[Miralda-Escud\'e(2000)]{escude}
Miralda-Escud\'e, J. 2000, in The First Stars, 
  MPA/ESO Workshop, eds. A. Weiss, T. G. Abel, \& V. Hill, 
  Springer-Cerlag, p.~242 

\bibitem[Moore et al.(2005)]{Moore}
Moore, B., Diemand, J., Madau, P., Zemp, M., \& Stadel, J. 2005, \mnras,
368, 563

\bibitem[Nakasato \& Nomoto(2003)]{nn03}
 Nakasato, N., \& Nomoto, K.\ 2003, \apj, 588, 842

\bibitem[Pieri, Martel, \& Grenon(2006)]{pmg06}
 Pieri, M., Martel, H., \& Grenon, C. 2006, submitted to ApJ
(astro-ph/0606423)

\bibitem[Preston, Shectman, \& Beers(1991)]{preston}
Preston, Shectman, Beers, 1991, ApJ, 375, 121

\bibitem[Scannapieco et al.(2005)]{stws05} 
Scannapieco, C., Tissera, P. B., White, S. D. M., \& Springel, V. 
2005, \mnras, 364, 552

\bibitem[Scannapieco et al.(2006)]{evan06} 
Scannapieco, E., Kawata, D., Brook, C. B., Schneider, R.,
Ferrara, A., \& Gibson, B. K. 2006, submitted to ApJ (Paper I)


\bibitem[Steinmetz \& Muller(1995)]{sm05}
Steinmetz, M., \& Muller, E. 1995, \mnras, 276, 549

\bibitem[Susa \& Umemura(2006)]{susa}
Susa, H., \& Umemura, M. 2006, submitted to ApJ (astro-ph/0604423)

\bibitem[Tegmark et al.(1997)]{tegmark} 
Tegmark, M., Silk, J., Rees, M. J., Blanchard, A., Abel, T., 
\& Palla, F. 1997, \apj, 474, 1


\bibitem[White \& Springel(2000)]{white}
White, S. D. M., \& Springel, V. 2000, in The First Stars, 
  MPA/ESO Workshop, eds. A. Weiss, T. G. Abel, \& V. Hill, Springer, p. 327 


\bibitem[Yoshida et al.(2003)]{yoshida}
Yoshida, N., Abel, T., Hernquist, L., \& Sugiyama, N. 2003, \apj, 593, 645

\bibitem[Yoshida (2006)]{yosh06}
 Yoshida,  N.  2006 NewAR,  50,  19

\end{thebibliography}
\end{document}